\documentclass[conference]{IEEEtran}
\usepackage{cite}
\usepackage{amsmath,amssymb,amsfonts}
\usepackage{algorithm}
\usepackage{algorithmic}
\usepackage{graphicx}
\usepackage{textcomp}
\usepackage{multirow}
\usepackage{listings} 
\usepackage{balance}
\usepackage{subfiles}
\usepackage{float}
\usepackage{url}
\def\BibTeX{{\rm B\kern-.05em{\sc i\kern-.025em b}\kern-.08em
    T\kern-.1667em\lower.7ex\hbox{E}\kern-.125emX}}

\begin{document}

\title{Unveiling Exception Handling Guidelines Adopted by Java Developers}

\author{\IEEEauthorblockN{
Hugo Melo, Roberta Coelho}
\IEEEauthorblockA{Federal University of Rio Grande do Norte \\
Natal, Brazil \\
hugofm@ppgsc.ufrn.br, roberta@dimap.ufrn.br}
\and
\IEEEauthorblockN{Christoph Treude}
\IEEEauthorblockA{University of Adelaide\\
Adelaide, Australia \\
christoph.treude@adelaide.edu.au}}


\maketitle

\begin{abstract}
Despite being an old language feature, Java exception handling code is one of the least understood parts of many systems. Several studies have analyzed the characteristics of exception handling code, trying to identify common practices or even link such practices to software bugs. Few works, however, have investigated exception handling issues from the point of view of developers. None of the works have focused on discovering exception handling guidelines adopted by current systems -- which are likely to be a driver of common practices. In this work, we conducted a qualitative study based on semi-structured interviews and a survey whose goal was to investigate the guidelines that are (or should be) followed by developers in their projects. Initially, we conducted semi-structured interviews with seven experienced developers, which were used to inform the design of a survey targeting a broader group of Java developers (i.e., a group of active Java developers from top-starred projects on GitHub). We emailed 863 developers and received 98 valid answers. The study shows that exception handling guidelines usually exist (70\%) and are usually implicit and undocumented (54\%). Our study identifies 48 exception handling guidelines related to seven different categories. We also investigated how such guidelines are disseminated to the project team and how compliance between code and guidelines is verified; we could observe that according to more than half of respondents the guidelines are both disseminated and verified through code inspection or code review. Our findings provide software development teams with a means to improve exception handling guidelines based on insights from the state of practice of 87 software projects.

\end{abstract}

\begin{IEEEkeywords}
exception handling, exception handling guidelines, qualitative study, Java development
\end{IEEEkeywords}

\section{Introduction}

Current applications have to cope with an increasing number of abnormal computation states that arise as a consequence of faults in the application itself (e.g., access of null references), noisy user inputs or faults in underlying middleware or hardware.
Therefore, techniques for error detection and handling are not an optional add-on to current applications but a fundamental part of them. The exception handling mechanism~\cite{goodenough1975exception}, embedded in most modern programming languages, is one of the most used techniques for detecting and recovering from such exceptional conditions. 

In Java, approximately 10\% of the source code is dedicated to exception handling and signaling~\cite{cabral2008case}. However, some studies have found that exception handling code is not only difficult to implement~\cite{robillard2003static} but also one of the least understood parts of the system~\cite{shah2008developers}. 

On the one hand, due to this lack of understanding regarding the exception handling behavior developers may think that by just sprinkling the code with catch-blocks in all places where exceptions may potentially be thrown, they are adequately dealing with the exceptional conditions of a system~\cite{mandrioli1992advances}. It may turn exception handling code into a generalized ``goto'' mechanism, making the program more complex and less reliable~\cite{mandrioli1992advances}.   

On the other hand, some guidelines have been proposed on how to use Java exceptions~\cite{mandrioli1992advances, gosling2000java, wirfs2006toward, bloch2008effective, coelho2018catalogue}. Such practices propose ways to implement the exception handling code in Java. 
However, there is a lack of empirical studies that try to investigate the motivations and decisions behind Java exception handling code and how these decisions can impact software development.

In this work, we present a qualitative study whose goal is to discover which guidelines have been used by Java developers to guide the development of exception handling code. This work aims at investigating the following research questions: \textit{Which exception handling guidelines are being used by Java projects? How are such guidelines disseminated among project members? How is the compliance between such guidelines and the code checked?}

Firstly, we performed semi-structured interviews with seven experienced developers from different companies, as an initial investigation into the exception handling guidelines being used. Based on the interview findings we designed a survey. The survey was sent to 863 GitHub developers and we received responses from 98 developers who were collaborating on 87 distinct projects. The survey contained both open and closed questions. The responses were then analyzed using the Grounded Theory techniques of open coding, axial coding and memoing \cite{charmaz2006constructing}. Our study leads to the following contributions:

\begin{itemize}
\item A characterization of the guidelines related to exception handling code that have been adopted in Java projects (Sec. \ref{subsec:RQ1-findings}).
\item An understanding of how such guidelines have been disseminated and checked against the code (Sec. \ref{subsec:RQ2-findings} and Sec. \ref{subsec:RQ3-findings}).
\item A set of practical implications related to the study findings (Sec. \ref{subsec:implications}).
\end{itemize}

\section{Background}

\subsection{Java Exception Handling Model}

In Java, exceptions are represented according to a class hierarchy, in which every exception is an instance of the Throwable class. Java exceptions can be of one of three kinds: checked exception (extends \texttt {Exception}), runtime exception (extends \texttt {RuntimeException})~\cite{gosling2000java}, and errors (extends \texttt {Error}). Checked exceptions must be declared in the method's \emph{exception interface} (i.e., the list of exceptions that a method might raise during its execution) and the compiler statically checks if appropriate handlers are provided within the system. Runtime exceptions are also known as ``unchecked exceptions'', as they do not need to be specified in the method \emph{exception interface} and do not trigger any compile time checking. There is a long-lasting debate about the pros and cons of using such kinds \cite{javatutorial,stackoverflow,jenkov} for the user-defined exceptions. Finally, errors are  used by the JVM to represent resource restrictions and are also ``unchecked exceptions''.

An exception can be explicitly signaled using the throw statement or implicitly signaled by the runtime environment (e.g., \texttt{NullPointerException}, \texttt{OutOfMemoryError}). Once an exception is thrown, the runtime environment looks for the nearest enclosing exception handler (Java's try-catch block), and unwinds the execution stack if necessary. This search for the handler on the invocation stack aims at increasing software reusability since the invoker of an operation can handle the exception in a wider context~\cite{miller1997issues}. A common way of propagating exceptions in Java programs is through exception chaining~\cite{fu2007exception}, also called remapping. The exception remapping happens when one exception is caught and a different one is thrown; the new exception can wrap the exception originally caught or can be thrown without storing the caught exception. 

\subsection{Exception Handling Patterns and Practices}

Sets of good and bad practices on how to use Java exceptions have been documented, such as the ones documented by: Gosling~\cite{gosling2000java}, Wirfs-Brock~\cite{wirfs2006toward}, Bloch~\cite{bloch2008effective} and Adamson~\cite{adamson}. Moreover, some tools have been proposed to automatically identify bad practices related to the exception handling code, such as Robusta~\cite{robusta}, SpotBugs~\cite{spotbugs}, SonarLint~\cite{sonar} and PMD~\cite{pmd}. While some of the good and bad practices are agreed upon by a majority of developers (such as, Exception Swallowing should be avoided~\cite{bloch2008effective,robusta,adamson}), others depend on indivudual opinions of authors/developers (such as, User-defined exceptions should be checked~\cite{gosling2000java}). A compiled set of such practices can be found in~\cite{coelho2018catalogue}.

\section{Research Method}
In this section, we present the research questions and the research methods adopted to collect and analyze the data to respond to them. The research questions are as follows:
\begin{itemize}
 \item RQ1: Which exception handling guidelines are adopted by Java developers?
 \item RQ2: How are such guidelines being disseminated among team members?
 \item RQ3: How is the compliance between the source code and such guidelines checked?
\end{itemize}

To answer these questions, we conducted a survey-based study as described next.

\subsection{Preliminary Studies}

The first phase of this study was based on a set of semi-structured interviews. The questions asked during the interviews were derived from our research questions. We asked interviewees questions such as ``\textit{In the projects you worked on so far, was there any documentation, formal recommendation, or informal (spoken) recommendation about how exceptions should be used in the project?}''. Seven experienced developers from different companies were contacted and a set of semi-structured interviews were conducted remotely. The group of interviewees were selected  opportunistically, via networks of collaborators and colleagues. Each interview lasted from 40 to 80 minutes, and after that, the interviews were transcribed and analyzed. Although the data analysis did not reach the saturation point, these interviews were useful as a preliminary study to get initial insights into the exception handling guidelines being used in Java projects and helped us to design the initial version of a survey targeting a broader group of developers.

The second phase of the study was based on a survey targeting GitHub developers. We chose GitHub since it is the most popular and widely used project hosting site, which also provides some public email addresses of software developers working on these projects. We designed the first version of a questionnaire and sent it to 50 GitHub Java developers selected at random from our sample. This first survey had six respondents, but the responses were difficult to relate to the research questions. Based on this observation, we refined the questions and sent the survey to another 50 developers. We got 9 responses, all coherent with the questions, which led us to send the questionnaire to the remainder of the sample.

\subsection{Survey}
\label{subsec:survey}

In this section, we introduce our method for participant selection, the design of the questionnaire, and our data analysis process.

\subsubsection{Participants Selection}
\label{subsubsec:participants}

In this work, we selected a subset of GitHub developers, following the guidelines proposed by \cite{munaiah2017curating} and \cite{kalliamvakou2014promises}. We used the Java API provided by GitHub, and the search was performed from the 10th to 15th of March 2018. The search stopped when we reached 5000 GitHub repositories. The repository with the least number of stars had 24. On average the repositories had 1356 stars. The selection criteria are detailed below.
\begin{itemize}
    \item Repository Selection
    \begin{itemize}
        \item Ordered by the numbers of stars -- As stars are a strong indicator of project quality, and can also be used to filter out toy projects.
        \item Repositories created before 01/01/2018 -- As a way to exclude projects that are too young.
        \item Repositories whose last commit on \textit{master branch} was performed 30 days ago at most -- As a way to exclude projects without recent activity.
        \item Repositories that were not Android -- Since the exception handling code of projects using the Android framework has specific characteristics that differ from non-Android projects \cite{fan2018large}.
    \end{itemize}
    \item Developer Selection
    \begin{itemize}
        \item have performed at least 5 commits in the last month -- As a way to contact active developers.
        \item have committed changes to Java code -- As a way to select Java developers.
        \item have a public email address on GitHub.
    \end{itemize}
\end{itemize}

Some of the selection criteria may sound too strict such as selecting developers who have performed at least 5 commits in the last month.
However, we intended to contact active and highly engaged developers who could be more aware of the guidelines adopted by the project. We could find 4449 repositories according to the repository selection criteria, and after applying the developer selection criteria, we collected the email addresses of 863 developers. We then sent the survey to all of them and received 98 responses (11\% response rate).

\subsubsection{Questionnaire}
We designed the survey questions based on the guidelines proposed by \cite{kasunic2005designing} and on the insights provided by the preliminary studies. The questionnaire was composed of 10 questions, with a mix of open and closed questions. All questions were optional, and the closed questions also had the option ``I don't know'', as an alternative to avoid respondent frustration. Table \ref{table:survey} presents the survey questions. 

\begin{table*}[htbp]
\centering
\caption{Survey questions}
\label{table:survey}
\begin{tabular}{r l}
\hline\hline 
1 & How many years of experience do you have in software development? \\
2 & While developing, I dedicate part of my time to read, write, or think about exception handling.\\
\hline
3 & My project has rules which define what exceptions should be thrown by methods and classes. \\
4 & If yes, could you describe one of these rules? \\
5 & My project has rules which define what classes or layers should catch exceptions. \\
6 & If yes, could you describe one of these rules? \\
7 & My project has rules which define what actions should be taken after an exception is caught. \\
8 & If yes, could you give an example of handling action? \\
\hline
9 & How is such knowledge disseminated to other contributors? \\
10 & How do you verify that such rules/guidelines are being used correctly? \\ [1ex] 
\hline 
\end{tabular}
\end{table*}

The questionnaire started with a question in which we asked about the experience the respondent had in software development. The next question was about the frequency with which the developers dealt with exception handling code (read, write or think about exception handling code). The subjective nature of this question was intentional to measure the developer's feeling about exception handling code. 
The next six questions were about the rules/guidelines that should be adopted by his/her project regarding the development of the exception handling code -- in this work, we use rules and guidelines as synonyms.
Questions 3, 5 and 7 asked about the existence of guidelines. The respondent could answer that: (i) there were no guidelines; (ii) s/he did not know about them; (iii) there were documented and hence ``explicit'' guidelines; or (iv) there were ``implicit'' guidelines (i.e., developers know and use them, but they are not documented in any way, not even in code comments). If any implicit or explicit guideline exists, the respondent was asked to give examples in subsequent questions (4, 6 and 8).

The last two questions were related to how such guidelines were disseminated among project members and how the compliance between the code and the guidelines was checked. These were multiple choice questions, and the answer options were based on the insights gained from the semi-structured interviews. The respondent could also choose the option ``Other'' and include a different response.

There was a draw for one 50 dollar Amazon gift card to the developers participating in the survey.

\subsubsection{Data Analysis}

The survey had three open-ended questions for which the respondents could provide examples of guidelines related to exception signaling and handling. Overall, we received 155 responses to these open-ended questions from the 98 respondents (some respondents did not respond to all questions). We adopted techniques from Grounded Theory to analyze the answers to these questions \cite{charmaz2006constructing}. First, we performed open coding, during which two authors of this paper collaboratively coded all answers. The coding process took four sessions of three hours each. The goal of the open coding was to identify the guidelines adopted. As a result, 189 codes were obtained.

We then performed axial coding in which the guidelines were refined, compared and grouped according to their origin. Axial coding was performed by the same two coders -- 7 categories and 48 subcategories of codes emerged. The next step was memoing: Memos were written for each category and subcategory describing the guidelines. Although we describe the analysis process in sequential order, this process was performed iteratively, with feedback loops (e.g., this means that during axial coding or memoing we could revisit and refine the open codes previously defined).

\subsection{Demographics}

\begin{figure}
  \centering
  \includegraphics[scale=0.46]{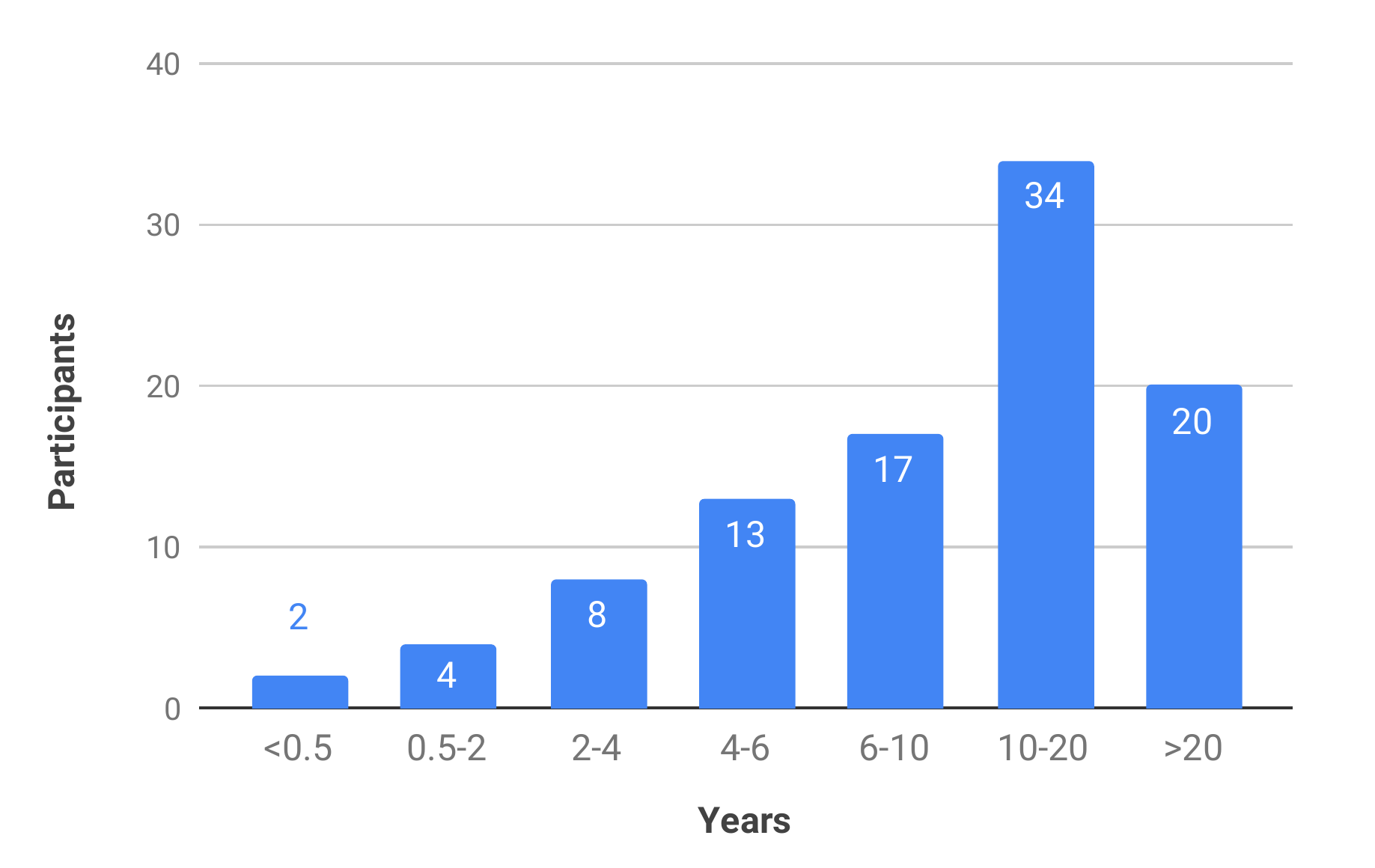}
  \caption{Experience in software development.}
  \label{fig:experience}
\end{figure}

In this section, we present the demographic information about the participants of the survey. Figure \ref{fig:experience} shows the number of years of software development experience per developer.

Only 6\% of the participants have two years of experience or less, and 72\% have six years or more.

\begin{figure}
  \includegraphics[scale=0.635]{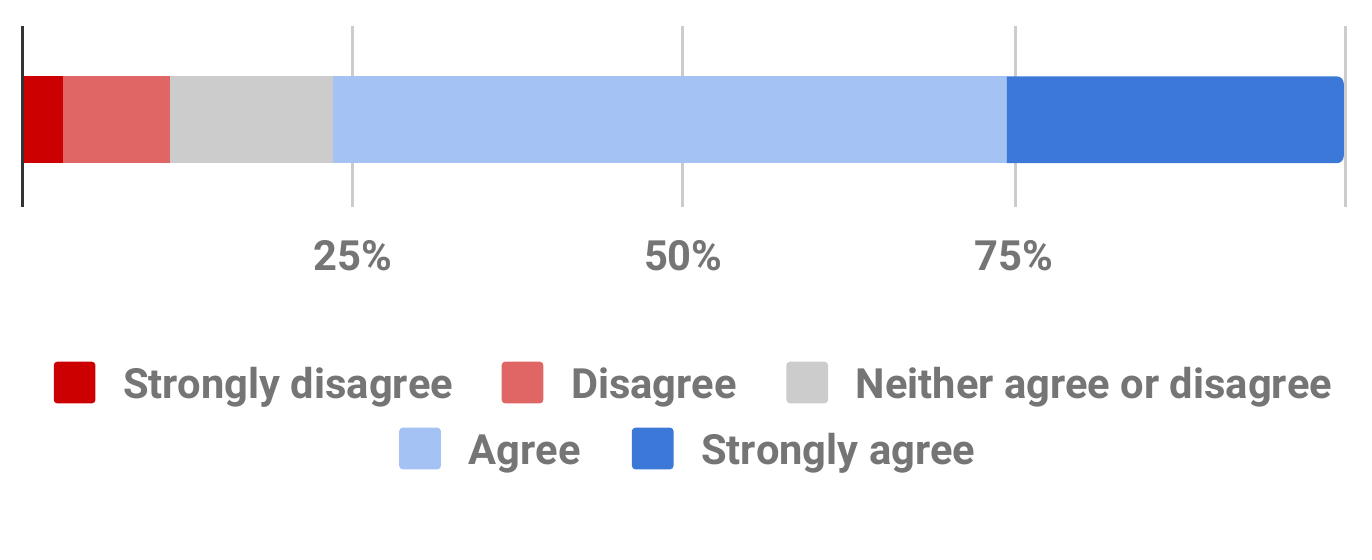}
  \caption{Agreement with ``While developing, I dedicate part of my time to read, write, or think about exception handling.''}
  \label{fig:dedication}
\end{figure}

We also investigated the dedication of participants regarding the development of exception handling code. Figure \ref{fig:dedication} shows the agreement of participants to the following statement: ``While developing, I dedicate part of my time to read, write, or think about exception handling.'' The minority of participants, 11\%, disagrees or strongly disagrees. The majority, 77\%, agree or strongly agree with the statement.

\section{Findings}

In this section, we present the survey findings. Section \ref{subsec:RQ1-findings} answers RQ1 by presenting the exception handling guidelines that emerged from our qualitative analysis. Sections \ref{subsec:RQ2-findings} and \ref{subsec:RQ3-findings} provide answers to RQ2 and RQ3.


\subsection{RQ1: Which exception handling guidelines are adopted by Java developers?}
\label{subsec:RQ1-findings}

In the questionnaire, we asked the developers about the existence of guidelines for signaling exceptions, for catching exceptions, and for  performing the handling actions (i.e., the actions performed after the exceptions are caught). There were four possible answers: 
(i) there were no guidelines; (ii) s/he did not know about them; (iii) there were documented and hence ``explicit'' guidelines; or (iv) there were ``implicit'' guidelines (i.e., developers know and use them, but they are not documented in any way, not even in code comments). Table \ref{table:existence} presents the responses.

\begin{table}[htbp]
\centering
\caption{Existence of guidelines related to exceptions}
\label{table:existence}
\begin{tabular}{l c c c}
\hline\hline 
\textbf{Option} & \textbf{Signaling} & \textbf{Catching} & \textbf{Handling} \\
\hline \\ [-1.5ex]
\textbf{No}             & 29 (30\%) & 43 (44\%) & 40 (41\%) \\
\textbf{Yes - implicit} & 53 (54\%) & 39 (40\%) & 38 (39\%) \\
\textbf{Yes - explicit} & 16 (16\%) & 12 (12\%) & 19 (19\%) \\
\textbf{Don't know} & 0 (0\%) & 4 (4\%) & 1 (1\%) \\ [1ex] 

\hline 
\end{tabular}
\end{table}

Most of the responses revealed that the projects have exception handling rules (70\% for exception signaling, 51\% for catching, and 59\% for handling actions). Among the projects that have rules, most of these rules are implicit (54\% for signaling, 40\% for catching, and 39\% for handling). Fewer responses mentioned the existance of explicit (i.e., documented) rules (16\% for signaling, 12\% for catching, and 19\% for handling). Other responses revealed that a considerable amount of projects do not have rules for exception signaling (30\%), catching (44\%), or handling actions (41\%). Moreover, few participants were unable to answer (0\% for signaling, 4\% for catching, and 1\% for handling). Next, we detail the guidelines (mentioned by developers as either implicit or explicit) that emerged from our qualitative analysis based on open coding, axial coding and memoing as described before.



\begin{figure*}
    \centering
    \includegraphics[scale=0.43]{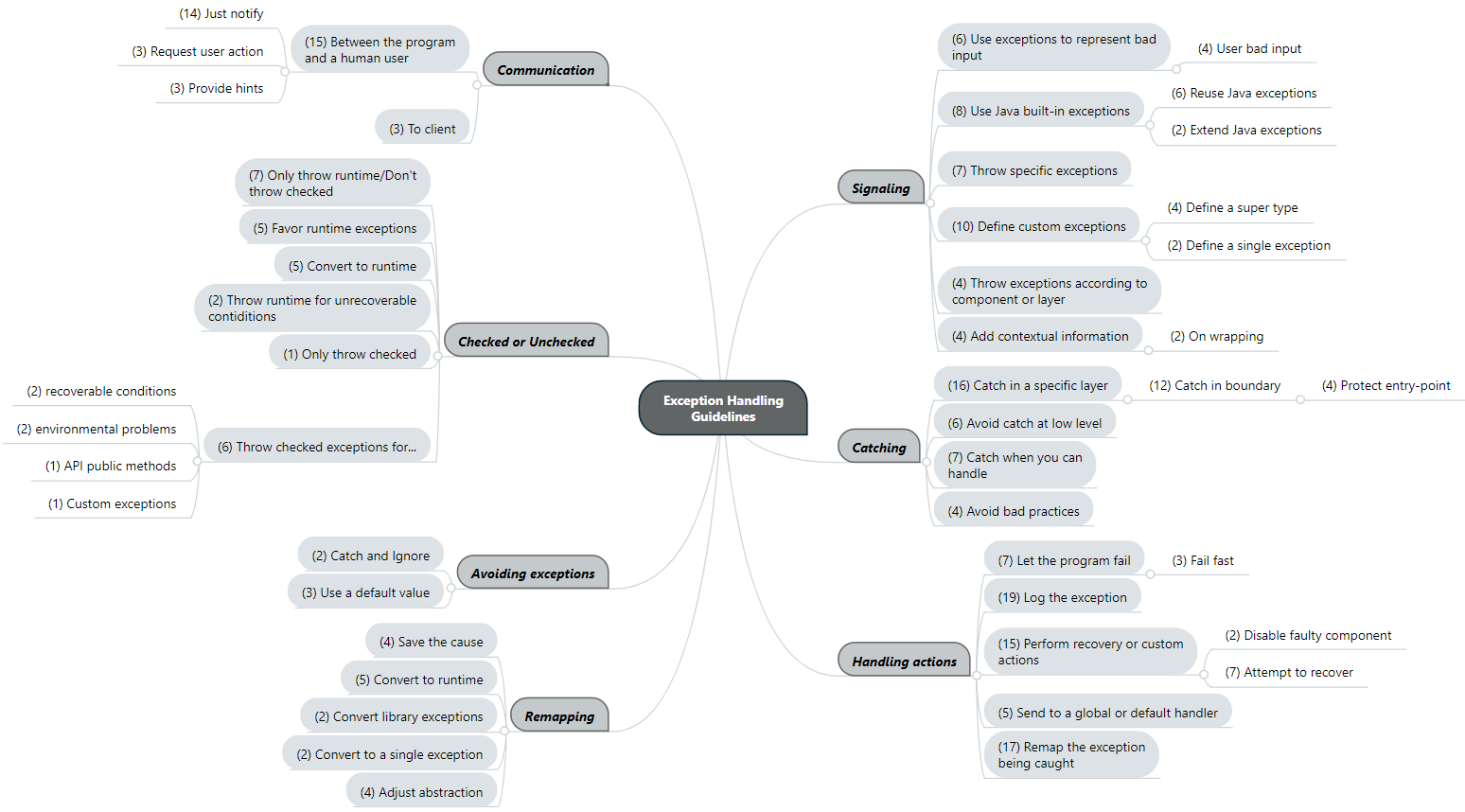}
    \caption{Exception Handling Guidelines.}
    \label{fig:mindmap}
\end{figure*}

Figure \ref{fig:mindmap} presents an overview of the guidelines that emerged. The numbers in the diagram represent the number of distinct participants that mentioned the given guideline, although occasionally one participant mentions the same guideline multiple times. The guidelines were divided in seven categories as follows:

\begin{enumerate}
\item signaling -- guidelines related to situations in which exceptions are thrown;
\item catching -- guidelines related to when and where exceptions should be caught;
\item handling actions -- guidelines related to the actions that should be performed when an exception is caught; 
\item checked or unchecked -- guidelines related to the exception types that should be adopted;
\item communication -- guidelines suggesting how exceptions can be used to communicate with the user; 
\item remapping -- guidelines detailing how and when exceptions should be remapped inside the system; and 
\item avoiding exceptions -- guidelines suggesting to avoid the use of exceptions.
\end{enumerate}

In the remainder of this section, we discuss the guidelines that emerged in this study and associate the participants who mentioned them; we also provide a selection of representative quotes to illustrate each finding. To ensure participants' anonymity, they are identified with the convention P\#.

\subsubsection{Signaling Guidelines}

Exceptional behavior starts when an exception is signaled with the \texttt{throw} clause. In this way, the type of the exception and its attributes, such as the message, will represent the failure that occurred and determines how this failure should be managed within the code. From our qualitative analysis, six core guidelines emerged for signaling exceptions: use Java built-in exceptions, define custom exceptions, throw specific exceptions, use exceptions to represent bad input, throw exceptions according to the component or layer, and add contextual information. The following sections explain each guideline.

\paragraph{Use Java built-in exceptions}

Java provides a wide variety of built-in exceptions. These exceptions are used by the Java core API, but can also be reused by developers. In our study, some (P28 P30 P48 P61 P73 P74 P86 P91) participants mentioned that the reuse of built-in exceptions was a guideline that should be adopted in their systems. The reuse of \texttt{IllegalArgumentException} and \texttt{IllegalStateException} was mentioned by participants (P30 P61 P73 P74 P91). The \texttt{IllegalArgumentException} is used to validate the input arguments of a method. The \texttt{IllegalStateException} validates the state of the operation or related objects. The reuse of \texttt{NullPointerException} was also cited by 1 participant (P28).

\begin{quote}
\textit{``I try to use Java's built in exception classes as much as I can before creating my own exception classes. For example, IllegalArgumentExceptions are thrown when a method parameter has an invalid value.  IllegalStateExceptions (less common) are thrown when there is something wrong with the internal state of an object.''} P61
\end{quote}

In addition to the direct reuse of Java types in signaling, some respondents (P48 P86) indicated that such built-in exceptions should be extended, creating specific custom exceptions.

\begin{quote}
\textit{``Use project specific sub class of generic Java Exceptions, i.e. $<$AnonymizedSQLException$>$ which extends java.sql.SQLException.''} P86
\end{quote}

\paragraph{Define custom exceptions}

The creation of custom exceptions was one of the guidelines mentioned by the participants (P8 P13 P35 P49 P56 P66 P74 P76 P86 P91). The answers indicated that custom exceptions could be used in different ways in a project. One way is to have a custom exception to be a super type of project exceptions (P8 P35 P74 P76).

\begin{quote}
\textit{``All domain-specific exceptions are a subclass of a single special abstract exception.''} P76
\end{quote}

A second reason for defining custom exceptions was to use a single custom exception for the entire project (P49 P66). 

\begin{quote}
\textit{``... all exceptions are encapsulated in a customized exception...''} P66
\end{quote}

A third reason for defining custom exceptions was to create specific exceptions (P8 P74 P76).

\begin{quote}
\textit{``(Throw) As specific as possible, including custom exceptions if needed''} P13
\end{quote}

\paragraph{Throw specific exceptions}

Exceptions, just like any class, can be defined at different levels of abstraction. Types such as \texttt{Exception}, \texttt{RuntimeException}, or a custom type that represents the root of failures in a project are generic exceptions. Specific exceptions are the opposite. They represent a well-defined fault, such as \texttt{IllegalArgumentException}. Participants (P8 P12 P13 P56 P69 P74 P76) reported that the creation and use of specific exceptions were a guideline in their projects.
\begin{quote}
\textit{``A new exception class is created for every possible failure in the flow.''} P56
\end{quote}

\paragraph{Use exceptions to represent bad input}

Exceptions can represent failures of different origins and natures. Participants (P2 P28 P60 P71 P85 P96) reported creating and signaling exceptions to represent bad input. In most cases, the bad input was given by the user (P60 P71 P85 P96).
\begin{quote}
\textit{``Exceptions that are caused by the user providing bad data should be a subclass of UserException.''} P71
\end{quote}

\paragraph{Throw exceptions according to the component or layer}

Four participants (P14 P26 P37 P88) indicated that the exceptions must be thrown in accordance with the component or layer they are flowing through. Thus, when signaling or re-signaling an exception, the exception type must be related to the component signaling it.
\begin{quote}
\textit{``We have [a] well-defined set of exception[s] for each layer. For instance; $<$AnonymizedException$>$ should be thrown if there's an issue while compiling.''} P37
\end{quote}

\paragraph{Add contextual information}

Since exceptions are regular Java objects, they can contain attributes and methods. Four participants (P1 P22 P27 P91) mentioned that exceptions should include additional information about the context of the failure.

\begin{quote}
\textit{``If defining a custom checked exception, it should contain information relevant to the exceptional state it represents. Do not just `extend Exception'.''} P91
\end{quote}

Two participants (P1 P22) mentioned that such contextual information is added to the exception during exception wrapping.

\begin{quote}
\textit{``Catch, wrap, and rethrow when you can add contextual information between what [the] caller did and what was actually attempted.''} P22
\end{quote}

\subsubsection{Catch Guidelines}

An exception that was signaled flows through the program's execution stack until it is caught. In our study, we asked if there were rules defining which classes or layers should catch exceptions. Four guidelines emerged from our qualitative analysis of the responses, which will be explained next.

\paragraph{Catch in a specific layer}

Participants (P6 P9 P13 P17 P27 P32 P37 P38 P43 P46 P48 P49 P52 P56 P67 P69) indicated that there is a specific layer in which exceptions should be caught. In most cases (P6 P13 P17 P27 P37 P38 P43 P46 P48 P49 P52 P56) that layer is a boundary layer. The boundary, in this case, is the last layer, class, or method through which the exception flows before it escapes the scope of the project in question. We call boundary an entry-point, such as the main method, the run method of the Thread class or the public methods of an API that is responsible for communicating with the user.

\begin{quote}
\textit{``RuntimeExceptions should be handled only in the main app code.''} P6
\end{quote}

\begin{quote}
\textit{``The internal exceptions are handled by the request handler, the outmost class for an API.''} P27
\end{quote}

Among these 12 participants that mentioned that exceptions should be caught in a boundary layer, four of them (P38 P46 P49 P52) described a pattern for handling exceptions known as ``protect entry-point'' or ``Safety Net'' \cite{haase2002java}. To protect the entry-point, all method code must be encapsulated by a large try block. This block must be associated with a generic catch block, such as \texttt{Exception} or even \texttt{Throwable}, thus ensuring the catching of any exceptions that may still be flowing, preventing the program to crash in an unpredictable way. Because the catch is generic, and the context in which the catch occurs is generally too broad (the catch of the main method can catch exceptions from all program locations), the handling actions usually log the exception, notify the user that a failure has occurred (if there is a user), and perform cleaning actions. This set of handling actions is also known as ``graceful shutdown''. Although none of the responses explicitly use the phrase ``protect entry-point'', they describe what the pattern suggests, as in the following responses:

\begin{quote}
\textit{``A non-caught exception should be caught at the top-most level block and be managed or be logged or reported, the program should gracefully shut down.''} P38
\end{quote}

\begin{quote}
\textit{``Exceptions should be caught at the level they can be dealt with, sometimes this means they bubble up all the way to the start of an application which catches all exceptions, logs them and then exits with a failure code.''} P49
\end{quote}

\begin{quote}
\textit{``All thread entry points must catch and log throwable.''} P49.
\end{quote}

\paragraph{Avoid catching at low level}

Instead of indicating where the exception should be caught, some participants (P6 P14 P17 P27 P61 P70) reported that catching and handling exceptions should not occur in low-level classes or layers.

\begin{quote}
\textit{``Low-level code should propagate exceptions to higher level clients; error handling policy does not belong in low level code.''} P14
\end{quote}

\paragraph{Catch when you can handle}

Participants (P13 P33 P48 P61 P70 P92 P95) mentioned the catching of exceptions should only occur if a meaningful response can be given. Some of the participants (P61 P70) are specific in saying that the exception catching should happen if it is possible to send a message to the user.

\begin{quote}
\textit{``If the layer can respond in a useful way to the exception (e.g. displaying an error message to the user), then handle it.  Otherwise, pass it up the stack.''} P61
\end{quote}

\paragraph{Avoid bad practices}

Some known bad practices such as generic catches \cite{bloch2008effective} and exception swallowing \cite{bloch2008effective} were mentioned by some developers (P9 P34 P41 P80) as practices to be avoided. However, some developers accepted some bad practices such as exception swallowing given that comments were added to mention the reason why this was done.

\begin{quote}
\textit{``Never catch generic Exception.''} P9
\end{quote}

\begin{quote}
\textit{``Don't ignore them, but if you do add a comment why it's okay.''} P41
\end{quote}

\subsubsection{Handling Actions Guidelines}

When an exceptional condition is detected, and an exception is signaled, the ideal scenario is one in which the exception is caught, and the program manages to recover automatically, delivering what had been planned. In many cases, however, recovery needs to be assisted by the user, or recovery is by no means possible. We asked developers if there are rules that define what actions should be taken after an exception is caught. The themes that emerged were the following: (i) let the program fail, (ii) log the exception, (iii) perform recovery or custom actions, (iv) send to a global or default handler, and (v) remap the exception being caught (discussed in Section \ref{subsubsec:remapping}).

\paragraph{Let the program fail}

Some participants (P1 P9 P33 P48 P89 P91 P98) have indicated that letting the program fail or explicitly shutting down the program in an error state are ways to deal with a signaled exception. Some of them (P33 P91 P98) even mentioned the ``fail fast'' technique \cite{shore2004fail}, in which the program is finalized as soon as it identifies an exceptional condition, avoiding that the cause of the failure distances itself from its manifestation, and also avoiding side effects such as data corruption in a database. Allowing the program to fail does not occur in all cases, however. Sometimes (P33 P48) a recovery action is attempted, or the technique is used for some types of exceptions.

\begin{quote}
\textit{``Generally we log the exception with a custom message plus the stack trace of the exception and then either continue (if possible) or exit the application (if not).''} P48
\end{quote}

\paragraph{Log the exception}

The most common handling action mentioned by the participants (P3 P6 P9 P13 P14 P17 P29 P34 P43 P49 P56 P67 P69 P78 P80 P84 P91 P92 P95) was to log the exception, which means persisting at least the exception name, message, and stack trace. Depending on the type of the exception, especially if it is a custom exception, other data can also be persisted. Some participants (P5 P48 P61) mentioned an extra effort given to logging the exception, making the log easier to understand.

\begin{quote}
\textit{``In many cases, the stack trace should be logged, along with a short description of what was happening at the time.''} P61
\end{quote}

\paragraph{Perform recovery or custom actions}

Participants indicated that there are rules for program recovery (P9 P13 P33 P36 P38 P48 P56 P66 P69 P77 P89) or custom actions (P27 P32 P83 P95) if a failure occurs. Some of them (P9 P36) described that the component that presents the fault is identified and disabled, thus keeping the rest of the program running.

\begin{quote}
\textit{``Fail out the functionality that can't work in the error condition. Contain the error and allow the rest of the system and experience to succeed.''} P9
\end{quote}

Other participants (P13 P33 P38 P48 P56 P66 P69) mentioned that an attempt to recover is made.

\begin{quote}
\textit{``~`retrying logic' is also allowed to catch an exception and retry the number of times (if the exception is a transient error)...''} P56
\end{quote}

Other participants (P27 P32 P83 P95) described that there are custom actions, which are executed depending on the type of failure, but did not explain in detail what these actions do.

\begin{quote}
\textit{``Handle it properly and fail based on the exception.''} P95.
\end{quote}

\paragraph{Send to a global or default handler}

Some participants (P42 P52 P60 P65 P69) indicated that in their projects there is a class or component of global access that is responsible for handling the exceptions.

\begin{quote}
\textit{``An uncaught exception handler is used for each thread and thread pool.''} P52
\end{quote}

One of them (P52) mentioned the Java interface \texttt{UncaughtExceptionHandler}, which is used to handle exceptions that occur in threads. Two other participants (P60 P65) indicated that the global handler is used to deal with unexpected exceptions.

\begin{quote}
\textit{``Methods that should not throw but do throw report errors to a global exception handler.''} P65
\end{quote}

\subsubsection{Remapping Guidelines}
\label{subsubsec:remapping}

In Java, it is possible to signal exceptions in the catch block. Java allows three re-signaling strategies: (1) an exception is caught and re-signaled without modification; (2) the exception is caught, and another exception is signaled; and (3) the exception caught is wrapped into a new exception, which is then signaled. The term remapping (or chaining~\cite{fu2007exception}) is used for the last two cases. The guidelines that emerged from our qualitative study and were related to remapping were the following: save the cause, convert to unchecked, convert library exceptions, convert to a single exception, and adjust abstraction.

\paragraph{Save the cause}

Four participants (P6 P22 P71 P91) indicated that they should always wrap the caught exception in the new exception before re-signaling it, avoiding to lose the cause of the original fault.

\begin{quote}
\textit{``If an exception is caught and re-wrapped into another exception type, you should always attach the original exception to the cause so that the stack trace is not lost.''} P71
\end{quote}

\paragraph{Convert to unchecked}

Five participants (P6 P26 P38 P74 P98) reported that one guideline in their projects is to remap caught exceptions into unchecked types. Two of them (P6 P26) said this should be done after catching IOExceptions.

\begin{quote}
\textit{``Re-throw IOExceptions as RuntimeExceptions.''} P6
\end{quote}

One participant (P74) said this should be done for very rare or impossible exceptions, and another (P98) justified this by saying that this practice avoids method signature pollution caused by checked exceptions.

\begin{quote}
\textit{``If the exception is impossible or extemely rarely thrown by a method, I try/catch it and throw a RuntimeException if it does happen.''} P74
\end{quote}

\begin{quote}
\textit{``Wrap checked exceptions in RuntimeExceptions to avoid polluting signatures.''} P98
\end{quote}

\paragraph{Convert library exceptions}

Two participants (P14 P38) reported that exceptions from libraries should be wrapped after being caught.

\begin{quote}
\textit{``Low level classes should wrap library exceptions in class-specific [exceptions] for expressiveness.''} P14
\end{quote}

\paragraph{Convert to a single exception}

Two participants (P49 P66) reported that in their projects the exceptions are wrapped and re-signaled into a single exception. 

\begin{quote}
\textit{``Collect low level exceptions into a single high level exception.''} P49
\end{quote}

\paragraph{Adjust abstraction}

Four participants (P27 P32 P38 P91) reported that exception remapping is done to adjust the level of abstraction of the exception that is flowing. One participant (P38) said that library exceptions are converted to custom project exceptions.

\begin{quote}
\textit{``Third party library exceptions must be caught and processed either to be recovered with a default value or to be wrapped inside an unchecked exception from the program's domain.''} P38
\end{quote}

The other three mentioned that the remapping occurs in specific layers of their projects.

\begin{quote}
\textit{``Do not propagate checked exceptions from lower layers to upper layers, either handle them or throw a layer specific exception.''} P91
\end{quote}

\subsubsection{Communication Guidelines}

The use of the exception handling mechanism as a means of communication was observed in the answers to the three questions that asked about exception handling rules. Such communication can happen between the program and a human user or between the program and client code.

\paragraph{Between the program and a human user}

When communication is perfomed with the user, the program must, in some way, convert the exception object into a message or other information that the user can understand. How accurate and informative this communication should be, will depend on multiple factors. Most participants (P14 P17 P37 P46 P56 P60 P61 P66 P67 P70 P77 P85 P89 P96) have described simple ways of communication, with the intention of notifying the user that a fault has occurred, or giving a brief description of the fault.

\begin{quote}
\textit{``notify the user with a `user-friendly' message if user is present.''} P14
\end{quote}

There are cases (P60 P85 P96) where exception information is used to give detailed information and even provide hints on how to correct the error.

\begin{quote}
\textit{``The error reports what steps are needed to avoid this error.''} P96
\end{quote}

There are also situations (P60 P66 P77) where the program interacts with the user in the event of a failure, receiving a new input from the user or providing additional information about the failure.

\begin{quote}
\textit{``Exceptions that we do not know how to handle (unexpected) are presented on the User Interface in friendly manner, followed by a link on which the user can report further details about the problem that just happened.''} P60
\end{quote}

\paragraph{Between the program and client code}

Three participants (P26 P27 P66) stated that the exception object is sent to the client of their code. In these cases, these participants were developing libraries.

\begin{quote}
\textit{``These (internal exceptions) are then converted to a suitable exception before being sent to the client.''} P27
\end{quote}

\subsubsection{Avoid Exceptions}

Some participants (P9 P30 P38 P42 P78 P95) stated that they avoid using exceptions in their projects. One way to achieve this, which is described by P30, P38, and P78, is to return \texttt{null} or another default value, rather than signaling the exception.

\begin{quote}
\textit{``Instead of an exception, I prefer returning null (or better, a sensible default value like an empty string).''} P30
\end{quote}

Two participants (P9 P42) simply indicated that they swallowed the exceptions.

\begin{quote}
\textit{``If exceptions do happen, just catch it and continue running the program.''} P42
\end{quote}

Two other participants (P30 P95) said that there is a preference for preventing or handling failure, rather than signaling an exception.

\begin{quote}
\textit{``we tend not to re-throw exception but to catch them.''} P95
\end{quote}

\subsubsection{Checked or Unchecked guidelines}

Some of the guidelines mentioned by participants (P6 P7 P26 P32 P33 P38 P44 P46 P51 P56 P70 P74 P84 P87 P91 P92 P97) were related to the type of the exception that should be used (either checked or unchecked). Most of the participants preferred runtime (unchecked) exceptions over checked exceptions. Six participants (P38 P46 P51 P70 P92 P97) mentioned that in their systems developers should only throw runtime exceptions (or should not throw checked exceptions).
Only one participant (P84) mentioned that in his/her system developers should only throw checked exceptions. Five participants (P32 P33 P56 P87 P91) mentioned that developers should favor runtime exceptions over checked.

\begin{quote}
\textit{``No checked exceptions must be thrown.''} P38
\end{quote}

Five participants (P6 P26 P38 P74 P98) indicated that developers should convert checked exceptions into runtime exceptions. Some of them specified the situations: when dealing with IOExceptions (P26 P74); or when dealing with unrecoverable situations (P7 P44). Six participants mentioned that checked exceptions should be thrown in specific situations, such as for recoverable conditions (P91), environmental problems (P32), and exceptions signaled by APIs (P28).

\begin{quote}
\textit{``Throw checked exceptions only for exceptional states that the client code could/should recover from.''} P91
\end{quote}

Some developers (P97 P98) mentioned that by using runtime exceptions, the method signature does not get polluted as illustrated in the quote below.

\begin{quote}
\textit{``Wrap checked exceptions in RuntimeExceptions to avoid polluting signatures.''} P98
\end{quote}

\subsection{RQ2: How are such guidelines being disseminated among team members?}
\label{subsec:RQ2-findings}

\begin{figure}
  \includegraphics[scale=0.66]{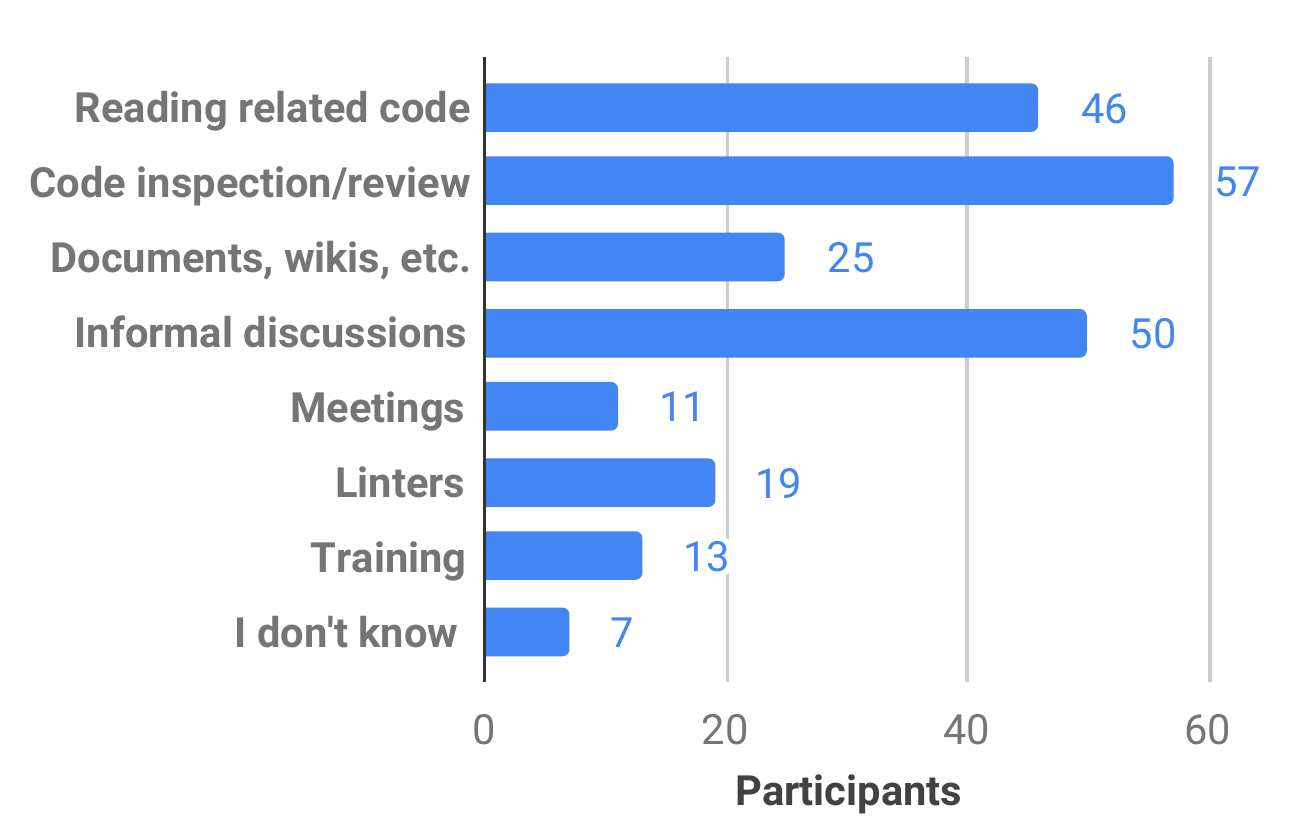}
  \caption{Strategies to disseminate Exception Handling Guidelines.}
  \label{fig:dissemination}
\end{figure}

One survey question asked about what strategies are used to disseminate guidelines among the development team (see Question 9 in Table \ref{table:survey}). This question got responses from 96 out of 98 participants (two of them did not answer this question), and respondents could select more than one option or add a new answer in an ``Other'' field -- but no respondent added an answer different from the presented ones. The most cited strategy was ``Code inspection / Peer Code Review'', showing up in 59\% of the answers; then ``Informal discussions'', in 52\%; and ``Reading related code from the same project'', in 48\% (see Figure \ref{fig:dissemination}). Similar results on exception handling guidelines dissemination were found by \cite{shah2010understanding}.

\subsection{RQ3: How is the compliance between the source code and such guidelines checked?}
\label{subsec:RQ3-findings}

\begin{figure}
  \includegraphics[scale=0.66]{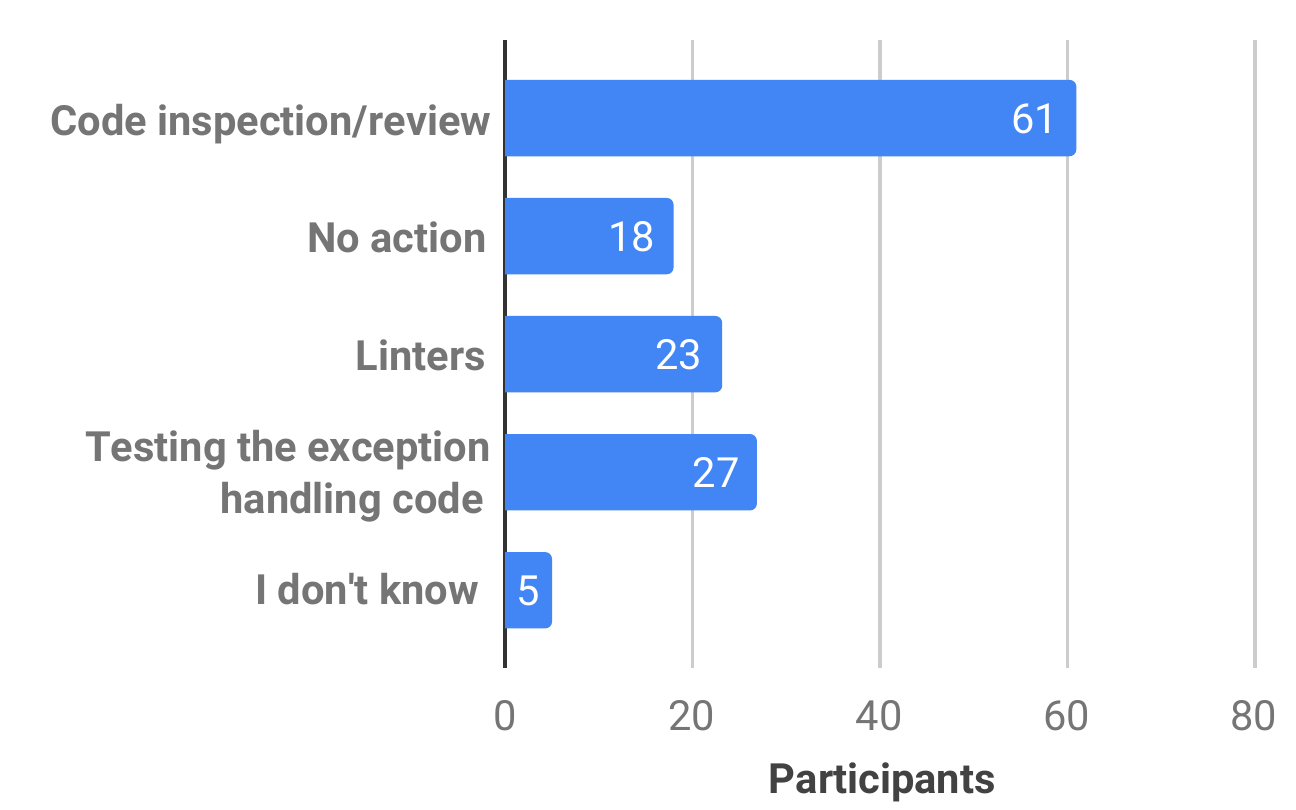}
  \caption{Strategies to verify compliance of Exception Handling Guidelines and source code.}
  \label{fig:verification}
\end{figure}

The question which asked about the strategies used to verify the compliance between the guidelines and the code was a multiple choice question (see Question 10 in Table \ref{table:survey}), for which respondents could select more than one option, and also add a new answer in an ``Other'' field -- but no respondent added an answer different from the presented ones. The most used strategy is ``Code inspection / Peer Code Review'', with 62\% of participant answers; then ``Testing the exception handling code'', with 28\%; and ``Linters'', with 23\%. Some participants (18\%) stated that no action is taken to verify compliance with rules. Figure \ref{fig:verification} shows the complete results. The number of developers who indicated that they perform specific tests to exercise the exceptional behavior is similar to the result found in \cite{ebert2015exploratory}.

\section{Discussions}

\subsection{Practical Implications} 
\label{subsec:implications}

\subsubsection{On the guidelines being used} The study revealed that most of the projects do have exception handling rules (70\% for exception signaling, 51\% for catching, and 59\% for handling actions). Such information is spread over several projects and is usually undocumented and implicitly defined. The study also revealed a considerable amount of projects that do not have guidelines for exception signaling (30\%), catching (44\%), or handling actions (41\%). The guidelines discussed in our study refer to the practices adopted by highly-starred GitHub projects. It was not our intent to judge if these practices follow the recommendations of literature or if they follow Java best practices for exception handling. Other studies\cite{barbosa2014categorizing,ebert2015exploratory,fan2018large,sena2016understanding,de2017studying} investigate the impact of some of the identified guidelines, but more work needs to be done on this subject.

\subsubsection{On the contribution to the debate about checked and unchecked exceptions}
Although the analyzed survey questions did not focus on issues relating to checked and unchecked exceptions, we could observe that several respondents mentioned guidelines related to the use of checked and unchecked exceptions inside the system. From such responses we could observe a tendency to favour runtime exceptions over checked exceptions. Most of the guidelines related to checked and unchecked exceptions advocated developers only to throw runtime exceptions or to favor runtime exceptions over checked exceptions. Only one respondent advocated the exclusive use of checked exceptions. The use of runtime exceptions is a characteristic of other languages (e.g., C++, C\#, Python, PHP, Javascript, VB .Net, Perl, Ruby), which may be influencing Java developers' preferences. However, further studies need to be conducted to investigate this issue. 

\subsubsection{On ways used to disseminate the guidelines} Most of the respondents mentioned that the guidelines were disseminated while performing code inspection or review, or just by reading related code or in informal meetings. This can indeed result in code that does not follow the intended guidelines. On the other hand, we have witnessed an increasing number of Linters that besides looking for non-conformances provide hints about the reasons for the non-conformity (e.g., SpotBugs \cite{spotbugs}, SonarQube \cite{sonar}). Such tools provide a way of disseminating best practices during coding. The previously mentioned tools are general purpose tools; in the context of exception handling code, there are also a few specific Linters (e.g., Robusta \cite{robusta}, Exception Policy Expert \cite{montenegro2018improving}). 

\subsubsection{On the compliance checking of guidelines} Most of the respondents mentioned that the compliance check between the guidelines and the code was performed by code inspection and review, which may result in a project that does not actually follow the exception handling guidelines. Hence, solutions aiming at automating the compliance checking of such guidelines (or part of them) and the code seem promising. Some work has been proposed in this direction \cite{abrantes2015specifying, barbosa2016enforcing, juarez2017preventing, montenegro2018improving}. However, in this work, the only guidelines that can be specified and automatically checked are the ones that can be expressed as: which elements should (or should not) throw and which elements should (our should not) catch exceptions. In our work, most of the mentioned guidelines were related to the way exceptions should be handled which cannot be expressed by the existing research work.

\subsection{Threats to Validity} 
\subsubsection{Projects} Mining GitHub data represents a threat since many projects on Github may be inactive and/or personal \cite{kalliamvakou2014promises}. We mitigate this threat by selecting mature and active projects, as presented in section \ref{subsubsec:participants}.

\subsubsection{Qualitative analysis} We chose methods from Grounded Theory to answer our research questions due to the exploratory nature of these questions. We perform collaborative coding in order to mitigate the bias of qualitative analysis.

\subsubsection{Guidelines} Although we achieved saturation when analyzing the survey responses (during collaborative coding), we cannot claim that we could get all possible guidelines adopted by GitHub developers since the general population on GitHub might have different characteristics and opinions.

\subsubsection{Project vs personal guidelines} Although we designed the questions to explicitly ask about project's guidelines, we cannot guarantee that developers' responses are related to project guidelines instead of personal guidelines.

\subsubsection{Developers} We cannot claim that our results generalize to all GitHub developers nor to closed-source developers nor to the entire population of Java developers. Despite this, they are representative of real guidelines actually being used.
\section{Related Work}

\subsubsection{Qualitative studies on how developers deal with exception handling code} Previous studies have investigated how developers deal with the exception handling code of a system \cite{shah2008developers, shah2010understanding, ebert2015exploratory}.  Shah et al. \cite{shah2008developers} interviewed nine Java developers to gather an initial understanding of how they perceive exception handling and what methods they adopt to deal with exception handling constructs while implementing their systems. Their study revealed that developers tend to ignore the proper implementation of exception handling in the early releases of a system, postponing it to future releases, or until defects are found in the exception handling code. In a subsequent study, Shah et al. \cite{shah2010understanding} interviewed more experienced developers in order to contrast the viewpoints of experienced and inexperienced developers regarding exception handling. The authors observed that although experienced developers consider exception handling code indistinguishable or inseparable from the development of the program's main functionality, they recognize that they also adopted the ``ignore for now'' approach previously in their career. Their work focused on implementation issues of exception handling code, they did not investigate what are the handling guidelines used during development. In our study we conducted a set of interviews and a survey focusing on the exception handling guidelines that may support the implementation decisions regarding exception handling code. 

\subsubsection{Exception handling code analysis} Several works \cite{de2017studying, sinha2004automated, cabral2007exception, de2017revisiting, coelho2008assessing, jo2004uncaught, kery2016examining, nakshatri2016analysis, asaduzzaman2016developers, chen2009exception} analyzed Java source code or Java bytecode through Static Analysis to identify and classify patterns in the exception handling code. The analyzed data ranges from simple code metrics, such as number of throw statements and catch blocks to complex data-flow analysis results such as the number of exceptional flows. The number of projects analyzed ranges from four \cite{sinha2004automated} to hundreds of thousands \cite{kery2016examining, nakshatri2016analysis, asaduzzaman2016developers}. These works show how exception handling is used in practice, but do not investigate what decisions the developers took that led to those practices.
Some work \cite{sena2016understanding, fan2018large, ebert2015exploratory, de2017studying, barbosa2014categorizing, dePadua2018studying} examines the evolution of the exception handling code in Open Source projects to identify the major causes of Exceptions Handling code failures. These works identify bad smells in exception handling code, but do not investigate what decisions led to that exception handling code.


\section{Concluding Remarks}

In this paper, we reported on a qualitative study designed
to investigate which exception handling guidelines are being used by Java developers, how such guidelines are disseminated to the project team and how the compliance between the code and the guidelines is checked. We performed a semi-structured interview with seven experienced developers that guided the design of a questionnaire used to survey a group of GitHub developers. We defined the selection criteria to choose highly-starred projects and active developers. We then contacted 863 GitHub developers and obtained 98 responses. The questionnaire was composed of closed and open questions. The open questions were analyzed using Grounded Theory techniques (open coding, axial coding and memoing). 

The study shows that exception handling guidelines usually exist (70\% of the respondents reported at least one guideline) and such guidelines are usually implicit and undocumented (54\% of exception signaling guidelines, 40\% of exception catching guidelines, and 39\% of handling actions guidelines -- see Table \ref{table:existence}). Our study identifies 48 exception handling guidelines related to seven different categories. We also investigated how such guidelines are disseminated to the project team and how compliance between code and guidelines is verified; we could observe that according to more than half of respondents the guidelines are both disseminated and verified through code inspection or code review. Our findings provide software development teams with a means to improve exception handling guidelines based on insights from the state of practice of 87 software projects.


\nocite{*}
\balance
\bibliographystyle{IEEEtran}
\bibliography{references}

\begin{thebibliography}{10}
\providecommand{\url}[1]{#1}
\csname url@samestyle\endcsname
\providecommand{\newblock}{\relax}
\providecommand{\bibinfo}[2]{#2}
\providecommand{\BIBentrySTDinterwordspacing}{\spaceskip=0pt\relax}
\providecommand{\BIBentryALTinterwordstretchfactor}{4}
\providecommand{\BIBentryALTinterwordspacing}{\spaceskip=\fontdimen2\font plus
\BIBentryALTinterwordstretchfactor\fontdimen3\font minus
  \fontdimen4\font\relax}
\providecommand{\BIBforeignlanguage}[2]{{%
\expandafter\ifx\csname l@#1\endcsname\relax
\typeout{** WARNING: IEEEtran.bst: No hyphenation pattern has been}%
\typeout{** loaded for the language `#1'. Using the pattern for}%
\typeout{** the default language instead.}%
\else
\language=\csname l@#1\endcsname
\fi
#2}}
\providecommand{\BIBdecl}{\relax}
\BIBdecl

\bibitem{goodenough1975exception}
J.~B. Goodenough, ``Exception handling: issues and a proposed notation,''
  \emph{Communications of the ACM}, vol.~18, no.~12, pp. 683--696, 1975.

\bibitem{cabral2008case}
B.~Cabral and P.~Marques, ``A case for automatic exception handling,'' in
  \emph{Proceedings of the 2008 23rd IEEE/ACM International Conference on
  Automated Software Engineering}.\hskip 1em plus 0.5em minus 0.4em\relax IEEE
  Computer Society, 2008, pp. 403--406.

\bibitem{robillard2003static}
M.~P. Robillard and G.~C. Murphy, ``Static analysis to support the evolution of
  exception structure in object-oriented systems,'' \emph{ACM Transactions on
  Software Engineering and Methodology (TOSEM)}, vol.~12, no.~2, pp. 191--221,
  2003.

\bibitem{shah2008developers}
H.~Shah, C.~G{\"o}rg, and M.~J. Harrold, ``Why do developers neglect exception
  handling?'' in \emph{Proceedings of the 4th international workshop on
  Exception handling}.\hskip 1em plus 0.5em minus 0.4em\relax ACM, 2008, pp.
  62--68.

\bibitem{mandrioli1992advances}
D.~Mandrioli and B.~Meyer, \emph{Advances in object-oriented software
  engineering}.\hskip 1em plus 0.5em minus 0.4em\relax Prentice-Hall, Inc.,
  1992.

\bibitem{gosling2000java}
J.~Gosling, \emph{The Java language specification}.\hskip 1em plus 0.5em minus
  0.4em\relax Addison-Wesley Professional, 2000.

\bibitem{wirfs2006toward}
R.~J. Wirfs-Brock, ``Toward exception-handling best practices and patterns,''
  \emph{IEEE software}, vol.~23, no.~5, pp. 11--13, 2006.

\bibitem{bloch2008effective}
J.~Bloch, \emph{Effective java}.\hskip 1em plus 0.5em minus 0.4em\relax Pearson
  Education India, 2008.

\bibitem{coelho2018catalogue}
R.~Coelho, J.~Rocha, and H.~Melo, ``A catalogue of java exception handling bad
  smells and refactorings,'' in \emph{Proceedings of the 25th Conference on
  Pattern Languages of Programs}.\hskip 1em plus 0.5em minus 0.4em\relax The
  Hillside Group, 2018.

\bibitem{charmaz2006constructing}
K.~Charmaz, ``Constructing grounded theory: A practical guide through
  qualitative research,'' \emph{SagePublications Ltd, London}, 2006.

\bibitem{javatutorial}
``Unchecked exceptions: The controversy,''
  \url{http://docs.oracle.com/javase/tutorial/essential/exceptions/runtime.html},
  October 2018.

\bibitem{stackoverflow}
``Java: checked vs unchecked exception explanation,'' \url{
  http://stackoverflow.com/questions/6115896/java-checked-vs-unchecked-exception-explanationl},
  October 2018.

\bibitem{jenkov}
J.~Jenkov, ``Checked or unchecked exceptions?''
  \url{http://tutorials.jenkov.com/java-exception-handling/checked-or-unchecked-exceptions.html},
  October 2018.

\bibitem{miller1997issues}
R.~Miller and A.~Tripathi, ``Issues with exception handling in object-oriented
  systems,'' \emph{ECOOP'97—Object-Oriented Programming}, pp. 85--103, 1997.

\bibitem{fu2007exception}
C.~Fu and B.~G. Ryder, ``Exception-chain analysis: Revealing exception handling
  architecture in java server applications,'' in \emph{Proceedings of the 29th
  international conference on Software Engineering}.\hskip 1em plus 0.5em minus
  0.4em\relax IEEE Computer Society, 2007, pp. 230--239.

\bibitem{adamson}
C.~Adamson, ``Exception-handling antipatterns blog,''
  \url{https://community.oracle.com/docs/DOC-983543}, October 2018.

\bibitem{robusta}
``Robusta plug-in,''
  \url{https://marketplace.eclipse.org/content/robusta-eclipse-plugin}, October
  2018.

\bibitem{spotbugs}
``Spotbugs,'' \url{https://spotbugs.github.io/}, October 2018.

\bibitem{sonar}
``Sonarqube,'' \url{https://www.sonarqube.org/}, October 2018.

\bibitem{pmd}
``Pmd,'' \url{https://pmd.github.io/}, October 2018.

\bibitem{munaiah2017curating}
N.~Munaiah, S.~Kroh, C.~Cabrey, and M.~Nagappan, ``Curating github for
  engineered software projects,'' \emph{Empirical Software Engineering},
  vol.~22, no.~6, pp. 3219--3253, 2017.

\bibitem{kalliamvakou2014promises}
E.~Kalliamvakou, G.~Gousios, K.~Blincoe, L.~Singer, D.~M. German, and
  D.~Damian, ``The promises and perils of mining github,'' in \emph{Proceedings
  of the 11th working conference on mining software repositories}.\hskip 1em
  plus 0.5em minus 0.4em\relax ACM, 2014, pp. 92--101.

\bibitem{fan2018large}
L.~Fan, T.~Su, S.~Chen, G.~Meng, Y.~Liu, L.~Xu, G.~Pu, and Z.~Su, ``Large-scale
  analysis of framework-specific exceptions in android apps,'' \emph{arXiv
  preprint arXiv:1801.07009}, 2018.

\bibitem{kasunic2005designing}
M.~Kasunic, ``Designing an effective survey,'' Carnegie-Mellon University,
  Pittsburgh, PA, Tech. Rep., 2005.

\bibitem{haase2002java}
A.~Haase, ``Java idioms-exception handling.'' in \emph{EuroPLoP}, 2002, pp.
  41--70.

\bibitem{shore2004fail}
J.~Shore, ``Fail fast,'' \emph{IEEE Software}, no.~5, pp. 21--25, 2004.

\bibitem{shah2010understanding}
H.~Shah, C.~Gorg, and M.~J. Harrold, ``Understanding exception handling:
  Viewpoints of novices and experts,'' \emph{IEEE Transactions on Software
  Engineering}, vol.~36, no.~2, pp. 150--161, 2010.

\bibitem{ebert2015exploratory}
F.~Ebert, F.~Castor, and A.~Serebrenik, ``An exploratory study on exception
  handling bugs in java programs,'' \emph{Journal of Systems and Software},
  vol. 106, pp. 82--101, 2015.

\bibitem{barbosa2014categorizing}
E.~A. Barbosa, A.~Garcia, and S.~D.~J. Barbosa, ``Categorizing faults in
  exception handling: A study of open source projects,'' in \emph{Software
  Engineering (SBES), 2014 Brazilian Symposium on}.\hskip 1em plus 0.5em minus
  0.4em\relax IEEE, 2014, pp. 11--20.

\bibitem{sena2016understanding}
D.~Sena, R.~Coelho, U.~Kulesza, and R.~Bonif{\'a}cio, ``Understanding the
  exception handling strategies of java libraries: An empirical study,'' in
  \emph{Proceedings of the 13th International Conference on Mining Software
  Repositories}.\hskip 1em plus 0.5em minus 0.4em\relax ACM, 2016, pp.
  212--222.

\bibitem{de2017studying}
G.~B. De~P{\'a}dua and W.~Shang, ``Studying the prevalence of exception
  handling anti-patterns,'' in \emph{Program Comprehension (ICPC), 2017
  IEEE/ACM 25th International Conference on}.\hskip 1em plus 0.5em minus
  0.4em\relax IEEE, 2017, pp. 328--331.

\bibitem{montenegro2018improving}
T.~Montenegro, H.~Melo, R.~Coelho, and E.~Barbosa, ``Improving developers
  awareness of the exception handling policy,'' in \emph{2018 IEEE 25th
  International Conference on Software Analysis, Evolution and Reengineering
  (SANER)}.\hskip 1em plus 0.5em minus 0.4em\relax IEEE, 2018, pp. 413--422.

\bibitem{abrantes2015specifying}
J.~Abrantes and R.~Coelho, ``Specifying and dynamically monitoring the
  exception handling policy.'' in \emph{SEKE}, 2015, pp. 370--374.

\bibitem{barbosa2016enforcing}
E.~A. Barbosa, A.~Garcia, M.~P. Robillard, and B.~Jakobus, ``Enforcing
  exception handling policies with a domain-specific language,'' \emph{IEEE
  Transactions on Software Engineering}, vol.~42, no.~6, pp. 559--584, 2016.

\bibitem{juarez2017preventing}
L.~Juarez~Filho, L.~Rocha, R.~Andrade, and R.~Britto, ``Preventing erosion in
  exception handling design using static-architecture conformance checking,''
  in \emph{European Conference on Software Architecture}.\hskip 1em plus 0.5em
  minus 0.4em\relax Springer, 2017, pp. 67--83.

\bibitem{sinha2004automated}
S.~Sinha, A.~Orso, and M.~J. Harrold, ``Automated support for development,
  maintenance, and testing in the presence of implicit control flow,'' in
  \emph{Proceedings of the 26th International Conference on Software
  Engineering}.\hskip 1em plus 0.5em minus 0.4em\relax IEEE Computer Society,
  2004, pp. 336--345.

\bibitem{cabral2007exception}
B.~Cabral and P.~Marques, ``Exception handling: A field study in java and.
  net,'' in \emph{European Conference on Object-Oriented Programming}.\hskip
  1em plus 0.5em minus 0.4em\relax Springer, 2007, pp. 151--175.

\bibitem{de2017revisiting}
G.~B. de~P{\'a}dua and W.~Shang, ``Revisiting exception handling practices with
  exception flow analysis,'' in \emph{Source Code Analysis and Manipulation
  (SCAM), 2017 IEEE 17th International Working Conference on}.\hskip 1em plus
  0.5em minus 0.4em\relax IEEE, 2017, pp. 11--20.

\bibitem{coelho2008assessing}
R.~Coelho, A.~Rashid, A.~Garcia, F.~Ferrari, N.~Cacho, U.~Kulesza, A.~von Staa,
  and C.~Lucena, ``Assessing the impact of aspects on exception flows: An
  exploratory study,'' in \emph{European Conference on Object-Oriented
  Programming}.\hskip 1em plus 0.5em minus 0.4em\relax Springer, 2008, pp.
  207--234.

\bibitem{jo2004uncaught}
J.-W. Jo, B.-M. Chang, K.~Yi, and K.-M. Choe, ``An uncaught exception analysis
  for java,'' \emph{Journal of systems and software}, vol.~72, no.~1, pp.
  59--69, 2004.

\bibitem{kery2016examining}
M.~B. Kery, C.~Le~Goues, and B.~A. Myers, ``Examining programmer practices for
  locally handling exceptions,'' in \emph{Mining Software Repositories (MSR),
  2016 IEEE/ACM 13th Working Conference on}.\hskip 1em plus 0.5em minus
  0.4em\relax IEEE, 2016, pp. 484--487.

\bibitem{nakshatri2016analysis}
S.~Nakshatri, M.~Hegde, and S.~Thandra, ``Analysis of exception handling
  patterns in java projects: An empirical study,'' in \emph{Proceedings of the
  13th International Conference on Mining Software Repositories}.\hskip 1em
  plus 0.5em minus 0.4em\relax ACM, 2016, pp. 500--503.

\bibitem{asaduzzaman2016developers}
M.~Asaduzzaman, M.~Ahasanuzzaman, C.~K. Roy, and K.~A. Schneider, ``How
  developers use exception handling in java?'' in \emph{Proceedings of the 13th
  International Conference on Mining Software Repositories}.\hskip 1em plus
  0.5em minus 0.4em\relax ACM, 2016, pp. 516--519.

\bibitem{chen2009exception}
C.-T. Chen, Y.~C. Cheng, C.-Y. Hsieh, and I.-L. Wu, ``Exception handling
  refactorings: Directed by goals and driven by bug fixing,'' \emph{Journal of
  Systems and Software}, vol.~82, no.~2, pp. 333--345, 2009.

\bibitem{dePadua2018studying}
\BIBentryALTinterwordspacing
G.~B. de~P\'{a}dua and W.~Shang, ``Studying the relationship between exception
  handling practices and post-release defects,'' in \emph{Proceedings of the
  15th International Conference on Mining Software Repositories}, ser. MSR
  '18.\hskip 1em plus 0.5em minus 0.4em\relax New York, NY, USA: ACM, 2018, pp.
  564--575. [Online]. Available:
  \url{http://doi.acm.org/10.1145/3196398.3196435}
\BIBentrySTDinterwordspacing

\bibitem{castor2007extracting}
F.~Castor~Filho, A.~Garcia, and C.~M.~F. Rubira, ``Extracting error handling to
  aspects: A cookbook,'' in \emph{Software Maintenance, 2007. ICSM 2007. IEEE
  International Conference on}.\hskip 1em plus 0.5em minus 0.4em\relax IEEE,
  2007, pp. 134--143.

\bibitem{sinha1998analysis}
S.~Sinha and M.~J. Harrold, ``Analysis of programs with exception-handling
  constructs,'' in \emph{Software Maintenance, 1998. Proceedings.,
  International Conference on}.\hskip 1em plus 0.5em minus 0.4em\relax IEEE,
  1998, pp. 348--357.

\bibitem{cristian1985exception}
F.~Cristian, ``Exception handling and software fault tolerance,'' in
  \emph{Reliable Computer Systems}.\hskip 1em plus 0.5em minus 0.4em\relax
  Springer, 1985, pp. 154--172.

\bibitem{jakobus2015contrasting}
B.~Jakobus, E.~A. Barbosa, A.~Garcia, and C.~J.~P. de~Lucena, ``Contrasting
  exception handling code across languages: An experience report involving 50
  open source projects,'' in \emph{Software Reliability Engineering (ISSRE),
  2015 IEEE 26th International Symposium on}.\hskip 1em plus 0.5em minus
  0.4em\relax IEEE, 2015, pp. 183--193.

\bibitem{greiler2012test}
M.~Greiler, A.~van Deursen, and M.-A. Storey, ``Test confessions: a study of
  testing practices for plug-in systems,'' in \emph{2012 34th International
  Conference on Software Engineering (ICSE)}.\hskip 1em plus 0.5em minus
  0.4em\relax IEEE, 2012, pp. 244--254.

\bibitem{barbosa2012heuristic}
E.~A. Barbosa, A.~Garcia, and M.~Mezini, ``Heuristic strategies for
  recommendation of exception handling code,'' in \emph{Software Engineering
  (SBES), 2012 26th Brazilian Symposium on}.\hskip 1em plus 0.5em minus
  0.4em\relax IEEE, 2012, pp. 171--180.

\bibitem{rahman2014use}
M.~M. Rahman and C.~K. Roy, ``On the use of context in recommending exception
  handling code examples,'' in \emph{Source Code Analysis and Manipulation
  (SCAM), 2014 IEEE 14th International Working Conference on}.\hskip 1em plus
  0.5em minus 0.4em\relax IEEE, 2014, pp. 285--294.

\bibitem{robillard2000designing}
M.~P. Robillard and G.~C. Murphy, ``Designing robust java programs with
  exceptions,'' in \emph{ACM SIGSOFT Software Engineering Notes}, vol.~25,
  no.~6.\hskip 1em plus 0.5em minus 0.4em\relax ACM, 2000, pp. 2--10.

\bibitem{singer2014software}
L.~Singer, F.~Figueira~Filho, and M.-A. Storey, ``Software engineering at the
  speed of light: how developers stay current using twitter,'' in
  \emph{Proceedings of the 36th International Conference on Software
  Engineering}.\hskip 1em plus 0.5em minus 0.4em\relax ACM, 2014, pp. 211--221.

\bibitem{pham2013creating}
R.~Pham, L.~Singer, O.~Liskin, F.~Figueira~Filho, and K.~Schneider, ``Creating
  a shared understanding of testing culture on a social coding site,'' in
  \emph{Software Engineering (ICSE), 2013 35th International Conference
  on}.\hskip 1em plus 0.5em minus 0.4em\relax IEEE, 2013, pp. 112--121.

\bibitem{garcia2001comparative}
A.~F. Garcia, C.~M. Rubira, A.~Romanovsky, and J.~Xu, ``A comparative study of
  exception handling mechanisms for building dependable object-oriented
  software,'' \emph{Journal of systems and software}, vol.~59, no.~2, pp.
  197--222, 2001.

\bibitem{viega2000can}
J.~Viega and J.~Vuas, ``Can aspect-oriented programming lead to more reliable
  software?'' \emph{IEEE software}, vol.~17, no.~6, pp. 19--21, 2000.

\bibitem{chang2002visualization}
B.-M. Chang, J.-W. Jo, and S.~H. Her, ``Visualization of exception propagation
  for java using static analysis,'' in \emph{Source Code Analysis and
  Manipulation, 2002. Proceedings. Second IEEE International Workshop
  on}.\hskip 1em plus 0.5em minus 0.4em\relax IEEE, 2002, pp. 173--182.

\bibitem{knodel2007comparison}
J.~Knodel and D.~Popescu, ``A comparison of static architecture compliance
  checking approaches,'' in \emph{Software Architecture, 2007. WICSA'07. The
  Working IEEE/IFIP Conference on}.\hskip 1em plus 0.5em minus 0.4em\relax
  IEEE, 2007, pp. 12--12.

\bibitem{barbosa2017global}
E.~A. Barbosa and A.~Garcia, ``Global-aware recommendations for repairing
  violations in exception handling,'' \emph{IEEE Transactions on Software
  Engineering}, 2017.

\bibitem{charmaz2014constructing}
K.~Charmaz, \emph{Constructing grounded theory}.\hskip 1em plus 0.5em minus
  0.4em\relax Sage, 2014.

\end{thebibliography}

\end{document}